# Synthesis of Predictable Global NoC by Abutment in Synchoros VLSI Design

Jordi Altayó González*, Dimitrios Stathis, Ahmed Hemani
KTH Royal Institute of Technology, Electrum 229, 164 40 Kista, Stockholm, Sweden
{jordiag,stathis,hemani}@kth.se

***Abstract*— Synchoros VLSI design style has been proposed as an alternative to the standard cell best design style; the word synchoros is derived from the Greek word choros for space. Synchoricity discretises space with a virtual grid, the way synchronicity discretises time with clock ticks. SiLago (Silicon Lego) blocks are atomic synchoros building blocks like Lego bricks. SiLago blocks absorb all metal layer details, i.e., all wires, to enable composition by abutment of valid; valid in the sense of being technology design rules compliant, timing clean and OCV ruggedized. Effectively, composition by abutment eliminates logic and physical synthesis for the end user. Like Lego system, synchoricity does need a finite number of SiLago block types to cater to different types of designs. Global NoCs are important system level design components. In this paper, we show, how with a small library of SiLago blocks for global NoCs, it is possible to automatically synthesize arbitrary global NoCs of different types, dimensions, and topology. The synthesized global NoCs are not only valid VLSI designs, their cost metrics (area, latency, and energy) are known with post-layout accuracy in linear time. We argue that this is essential to be able to do chip-level design space exploration. We show how the abstract timing model of such global NoC SiLago blocks can be built and used to analyse the timing of global NoC links with post layout accuracy and in linear time. We validate this claim by subjecting the same VLSI designs of global NoC to commercial EDA's static timing analysis and show that the abstract timing analysis enabled by synchoros VLSI design gives same results as the commercial EDA tools.**

***Keywords* — *Coarse Grain Reconfigurable Architectures, Clock Tree Synthesis, VLSI design, SiLago***

## I. INTRODUCTION

Wires are scaling worse with technology, compared to computation and storage as shown in Fig. 1a. It is also well known that storage and interconnect dominate the cost metrics [1] as shown in Fig. 1b. This is because of the emergence of memory intensive applications especially in the ML/AI category [2]. These arguments suggest that interconnect cost should be factored in during the design space exploration (DSE) at all three levels of chip-design hierarchy: blocks, sub-system and chip. However, the interconnect cost is not known until physical design has happened. Many estimation techniques have been proposed to factor in the interconnect cost, especially for NoCs, however, these estimation techniques, especially at higher abstractions, face two challenges: i) the estimation accuracy is exponentially worse compared to the accuracy at lower abstractions and ii) the impact of interconnect cost on overall design cost becomes increasingly significant compared to the impact at lower abstractions. The abstractions from highest to lowest levels are: system, application, algorithms, RTL, logic/gate, physical design, see [3], for more details.

*Corresponding author

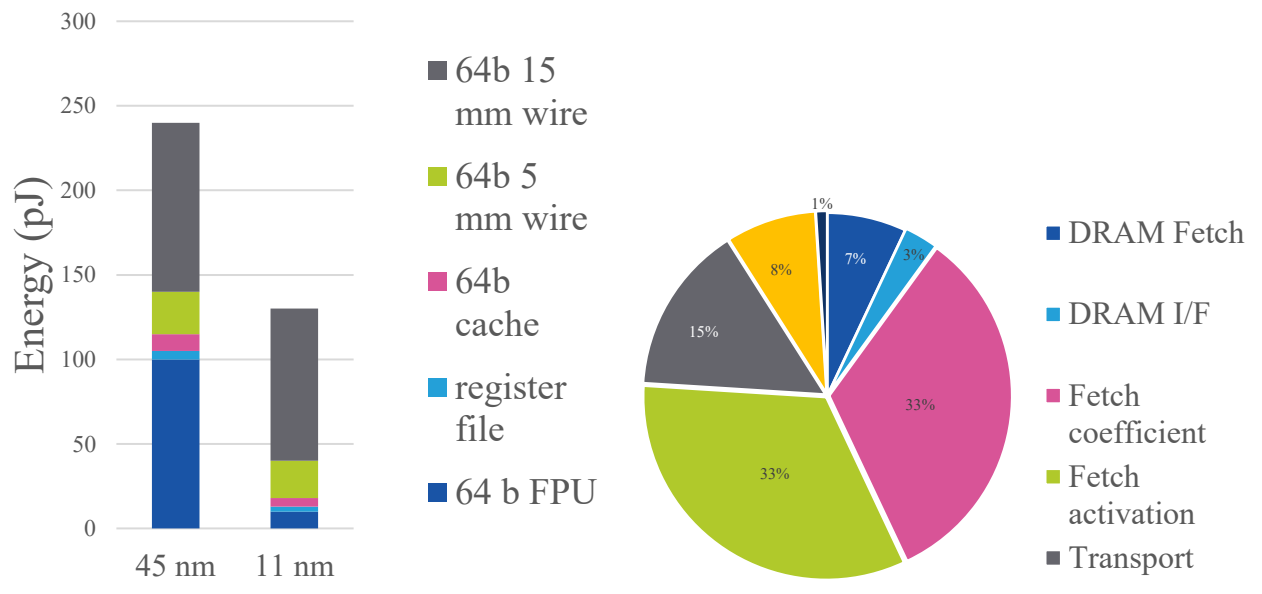

Figure 1. Breakdown of the energy consumption [1][2]

### A. *Problem statement*

To precisely define the problem being addressed in this paper and how the contributions of this paper are intended to be used, consider the abstract chip-level design space exploration (**DSE**) loop shown in Listing 1.

In the DSE loop, an application is mapped to a chip in terms of synchronous islands containing logic, storage, or both, that communicate with each other over lobal NoC (**GNoC**) and in general on latency insensitive basis. There are three key problems in the DSE loop:

a) Creating new solution (*create_new_solution* on line 3) in terms of deciding the number and content of the synchronous islands and the global NoC that meets the communication needs of these synchronous islands. These solutions are abstract, the VLSI implementation and post-layout area, latency, and energy are not known.
b) Implement the functionality mapped to synchronous islands as VLSI design and know their post-layout accurate costs. Floorplan the chip in terms of synchronous islands with space for GNoC.
c) Implement the VLSI design of GNoC according to the specification of GNoC created by step a) and the floorplan constraints created by step b)

We observe that the loop shown in Listing 1 is expected to deal with large designs of the order of 100s of million gates and the loop is expected to potentially evaluate millions of solutions. Synchoros VLSI design's post-layout accuracy and linear time complexity are essential for effective and efficient

Listing 1. Design space exploration example loop

```
1. cost_of_best_solution = infinity;
2. LOOP
3.   [Synchronous Islands, Global NOC] =
4.                      create_new_solution (Application);
5.     create_new_solution(application);
6.   cost_of_new_solution = cost of Synchronous Islands +
7.                       cost of Global NoC;
8.   IF cost_of_new_solution < cost_of_best_solution THEN
9.     cost_of_best_solution = cost_of_new_solution
10.    best_solution = new_solution
11.  END IF
```

DSE. The conventional state-of-the-art methods to estimate cost-metrics sacrifice accuracy for speed. The option to do logic and physical synthesis for millions of candidate solutions is obviously not scalable. Synthesis of GNoC in terms of SiLago blocks complements the research to improve predictability of NoCs at architectural and algorithmic level [4][5][6].

Composition of regional clock tree by abutment in a synchoros design framework has been elaborated [7]; this work has been presented at TAU 2021. A conceptual scheme to create power grid by abutment has also been presented [3]. The focus of this paper is on synthesis of GNoC by abutment.

We next introduce the essential concepts of synchoros VLSI design, as they are not widely known. This is essential to concretely define the contributions and for understanding the concepts and methods introduced in this paper.

## B. *Synchoros VLSI design*

Synchoricity is analogous to synchronicity. In synchronicity, time is uniformly discretised with clock ticks to simplify the temporal and logical composition. In synchoricity, space is uniformly discretised with a virtual grid to simplify the spatial and electrical composition that is technology compliant. Lego is a prime example of synchoros design objects; all Lego objects are integer multiples of a standard pitch and Lego studs are placed at these pitches to enable composition by abutment. This enables *infinite* variety of Lego structures to be built out of a *finite* set of Lego bricks type. Synchoros VLSI design style objects are called SiLago (Silicon Lego) blocks (**SB**). SBs, like their inspiration, the Lego bricks, are designed to compose large VLSI designs by abutment; no logic or physical synthesis is required. In this paper, we will introduce a set of GNoC SBs, that enables creation of arbitrary type and dimension of GNoC, simply by abutment, and with knowledge of post-layout accurate cost-metrics. We next introduce the three levels of hierarchical synchoros design objects: SBs, regions and chips.

SiLago blocks are the atomic building blocks and their salient properties are: a) they absorb all metal layer details, i.e., absorb all inter-SB wires, as part of their design; all implies functional wires like NoC, and infra-structural like clock, reset, power grid etc. b) bring out the inter-SiLago interconnects, like the Lego studs, at right place on the right metal layer. As a result, when *valid* SB neighbours [2] are placed on the grid, their interconnects align to create a larger, valid VLSI design, c) SiLago blocks export micro-architecture level operations in contrast to the boolean-level operations exported by the standard cells, d) SBs are characterized with post-layout data. Since, the intra-SB details and the inter-SB details are known and characterized with post-layout data, it is possible to create *abstract* models for a design composed in terms of SBs and know the cost-metrics with post-layout accuracy; see [8] for more details.

Regions are aggregates of SBs of the same type. In terms of the DSE loop in Listing 1, the reader can think of region instances as synchronous islands. For more details on regions, see [7]. Instances of region communicate with each other over GNoC and in general the communication can be latency insensitive. We have proposed using a variant of GALS protocol called GRLS [9].

[2] Every SB type can abut with a sub-set of other SB types. These are called *valid neighbours*.

Chips are aggregates of region instances or equivalently synchronous islands. Figure 2 shows a visual depiction of a synchoros VLSI Chip level floor plan. Multiple instances of different types of regions are shown; we remind that region instances are equivalent to synchronous islands in the DSE loop in List. 1. GNoC SBs are also shown creating the GNoC that connects the region-instances. Light pink shaded GNoC SBs are switchbox type of SBs, and the dark pink shaded SBs are wired, buffered, or registered SBs. Light shaded SBs in region instances are Network Interface Units (NIUs) that connect the region instance to the GNoC.

## C. *Contributions of this paper*

1. A set of SBs types to create an arbitrary GNoC and distribute a global clock tree (**GCT**) along with it.
2. Characterization and timing model of the GNoC SBs to enable linear time and post-layout accurate timing analysis of arbitrary GNoC composed from such SBs.
3. Synthesis algorithm to select the optimal mix of GNoC SBs that fulfils the abstract GNoC specifications created by the *create_new_solution* in the DSE loop in Listing 1.
4. Validation of the claim of linear time and post layout accuracy of the cost-metrics of GNoC composed by abutment of SBs.

The rest of the paper is organized as follows. In section II, we introduce the GNoC SBs, their characterization, timing model and synthesis of GNoC in terms of the SBs. In section III, we present the experimental results and validate the claim; see point 4 above. In section IV, we review the state-of-the-art and differentiate the proposed method and justify it. Finally, in section V, we draw conclusions and present a list of enhancements that we are working on to go beyond the results presented in this paper.

# II. Analysis and Synthesis of Synchoros GNoC

In this section, we introduce a set of GNoC SB types that are sufficient to implement arbitrary type and dimensions of GNOC. Each type of GNoC SB has sub-types to cater to variations in dimensions (width) and functionality, i.e., different types of routers with varying dimensions of storage for packets, commands, status etc. We also introduce the timing model of the SBs to enable linear time, post layout accurate timing analysis that we call higher abstraction static timing analysis or **HASTA**. Finally, we present a synthesis algorithm to synthesize arbitrary GNoC in terms of an optimal set of GNoC SB instances.

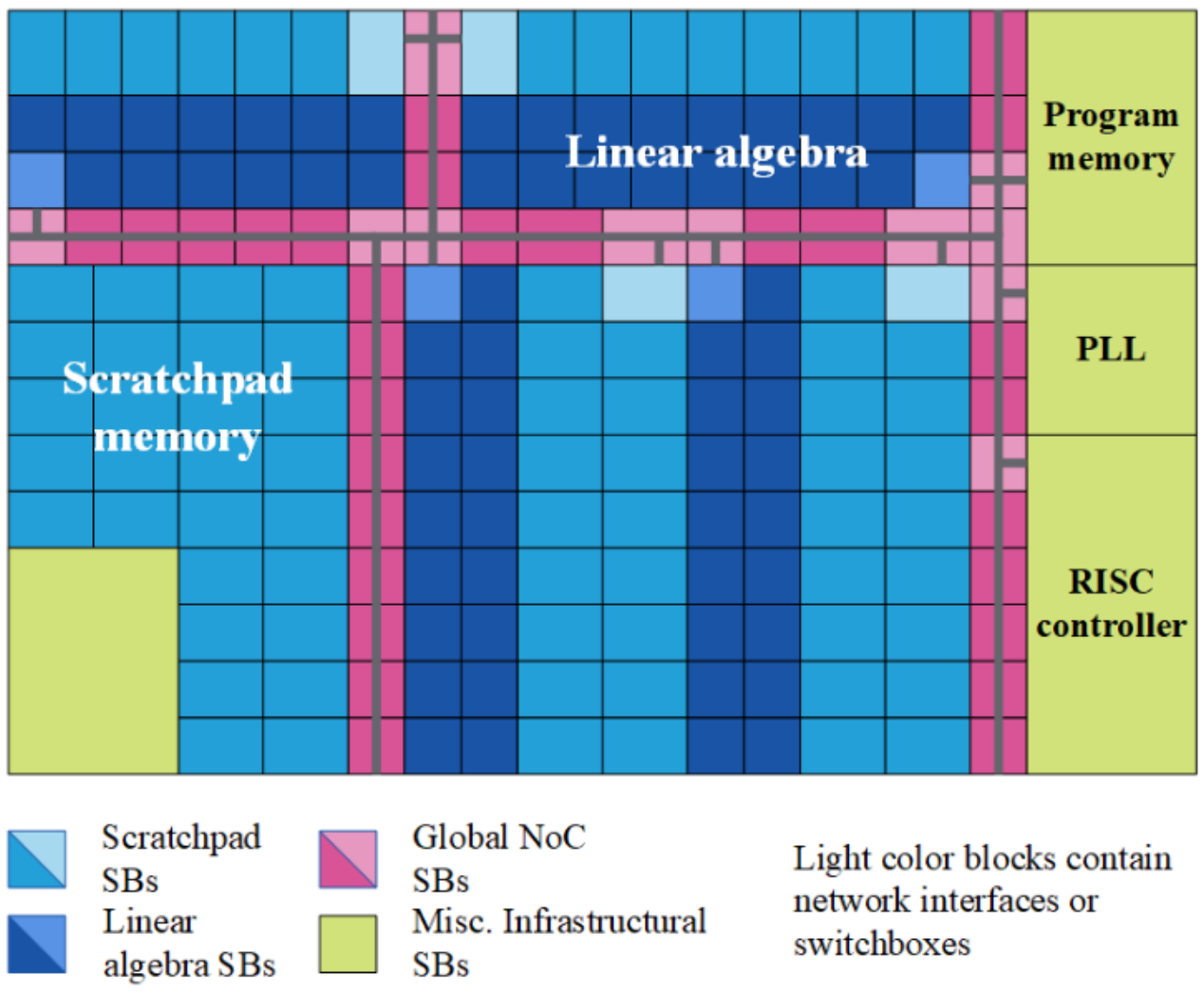

Figure 2. SiLago floorplan example.

Besides GNoC fragments, the SBs also absorb fragments of the GCT. As a result, both GNoC and GCT gets created by abutment of these SBs. For brevity, we will only mention GNoC but also imply GCT as well. We mention GCT, only when arguments for GNoC, do not apply to GCT.

### *A. Abutable GNoC SiLago Blocks (SBs)*

There are three types GNoC SBs to build NoC wires and another type for switchbox and router logic. These blocks propagate NoC wires as well as a global clock. Some of these blocks also use the global clock besides propagating it. The logical layout of the GNoC in terms of these SiLago blocks can be expressed as a simple grammar presented in Listing 2. Every GNoC instance in a synchoros VLSI design is a sentence that fulfils this simple grammar. Every GNoC SB sub-type is characterized, and this data is then used for the timing model that is used by HASTA. Each sub-type of GNoC SB is also made abutment ready and the way wires are laid out ensures that technology design rules are also fulfilled in the design created by abutment. We next elaborate the function, design, and dimensions of these GNoC SBs.

*Wire (`W`) GNoC SB:* The plain wire SB, represented by token `W` in Listing 2 and shown in Fig. 3.1, is the simplest and used to create wires of length that allows the slew rate to deteriorate within the lower bound of slew-rate spread that the technology rule allows. Note that the drivers of these wires and their strength is known because the driver is also part of some pre-characterized SB that drives these wire SBs. In general, the need to buffer the GNoC wires is similar, with a small spread in slew rate variations. However, the GCT wire, shown as red lines in Fig. 3, could have larger variations because it is expected to be more heavily loaded. For this reason, the wire (`W`) GNoC SiLago blocks (Fig 3.1) come in two sub-types: one in which the GCT fragment is buffered and the second in which, its non-buffered, as shown in Fig 3.1.

*Buffer (`B`) GNoC SB:* The buffered wire SB, represented by token `B` in the Listing 2 and shown in Fig. 3.2 is used to restore the slew rate to a healthy value within the slew rate spread that the technology rule prescribes. In this SB, all wires are buffered.

*Register (`R`) GNoC SB:* Finally, the registered wire SB, shown as token `R` in Listing 2 and depicted Fig. 3.3, is used to pipeline the GNoC wires, should the propagation delay through the wire and buffered SBs become longer than the global-clock period. It should be noted that this applies only to the GNoC wires but not to the GCT. For the GCT wire, it is essential to have the minimum distance between buffers that is less than half the clock period of the maximum intended global clock frequency.

*Switch/Router (`S`) GNoC SB:* The GNoC switch/router, denoted as `S` in Listing 2, and conceptually depicted in Fig. 3.4. Depending upon the type of Switch/Router, the dimension of storage etc. there will multiple sub-types of `S` type GNoC SBs. However, all of them would be synchoros and composable with other GNoC SBs by abutment.

*Network Interface Unit (`N`) GNoC SB:* The `N` type of GNoC SB is placed in *region* instances, see section I.B. N SBs are not shown in Fig. 3 because their instantiation is part of region instance synthesis that is not the focus of this paper.

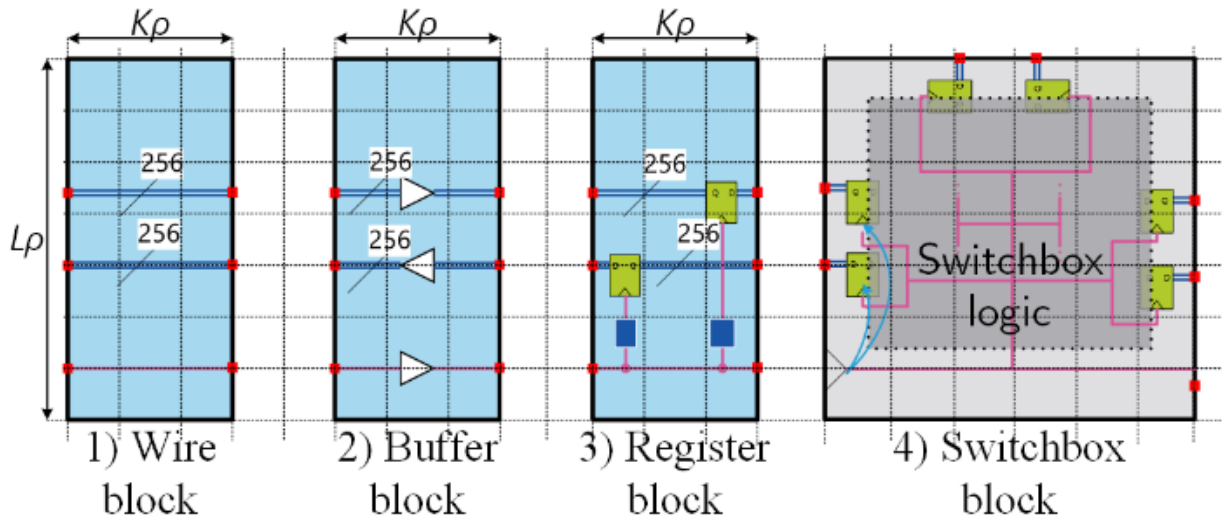

Figure 3. Abutable GNOC SBs used in the GNoC edges

*Synchoricity of GNoC SBs* constraints three aspects of VLSI design of these SBs. i) Width varies according to the content; width of `W` SBs would be least and `S` SB the maximum. All widths will be an integer multiple of the pitch of the virtual grid as shown in Fig. 3. ii) Height is the same for SBs that carries the same number and type of wires and is also an integer multiple of the pitch of the virtual grid. iii) SBs that carries the same number and type of wires form a family of GNoC SB sub-types that are valid neighbours, and they all bring out their wires on the periphery at right place and right metal layer to enable composition by abutment. The layout of the wires is also controlled to comply with the technology design rules.

### *B. Characterization, Timing models and Analysis*

In this section, we elaborate how these SBs are characterised, and the characterization data used to create timing models that predict post-layout accurate timing in linear time and also how these models are made PVT and OCV rugged.

*Characterization*

As mentioned in the section I.B, a synchoros VLSI design is composed by abutting the pre-characterized, hardened SBs. This implies that all *instances* of a specific SB sub-type will be identical in their VLSI design. For this reason, characterization is a onetime time effort for each SB sub-type, as is the case with characterizing standard cells.

Characterization distinguishes between two kinds of SBs: the ones with active logic (`S`, `R` and `B`) and the others that are passive with plain wires (`W`). The characterization of the passive wire SBs is simple, it encapsulates the length of the wire that is an integer multiple of the pitch of the synchoros VLSI design. This length of wire translates into capacitive load for the active SBs that drives the plain wire (`W`) SBs.

Every GNoC timing path is divided into segments. Each segment starts and ends with an active [`R`|`S`|`B`] SB with up to *K* intervening passive `W` SBs; *K* depends on the technology and the width of the SBs. Since there are three active SB types, there are 9 possible combinations of source and

Listing 2. The GNoC floorplan grammar

```
// Tokens
S: Switch Box/Router
W: Plain Wires
B: Buffered Wires
R: Registered Wires

// GNOC grammar
GNOC: S Wires GNOC |
      S
Wires: {W, B, R}+
```

destination to create 9 types of segments. For each segment type, one characterization table is built with two dimensions: range (*L*) of input slew rates from low to high and the load corresponding to the number (*0* to *K-1*) of intervening `W` SBs. In our experiments, we have used *K=L=10*. This implies 9 tables of size 10×10 cells. Since the SBs are modest in their complexity, invoking static timing analysis (**STA)** 900 times takes only 1 or 2 hours, and even that is a onetime effort.

While each segment instance of GNoC has 900 possible variants, a GNoC will be a cascade of 100s of such segments making the design space of GNoCs virtually infinite. *However, it is possible to analyse **any instance** from among the infinite variants of GNoC in linear time and with post-layout accuracy. This is because, if there are N segments of a GNoC, HASTA just needs to do N look ups in one of the 9 2-dimensional tables.*

To use the two-dimensional table, HASTA first decides which of the 9 table to use depending on the source and destination active block combination. The column is decided by the intervening number of wire blocks. The row is selected based on the input slew rate. If it is not an exact match, then there are two options. One is to take a pessimistic view and use the next row with worse slew rate for setup-checks and better slew rate for hold checks. The second option is to consider the two rows corresponding to the two closest slew rates are selected and perform linear interpolation. A third option is to increase *L* to improve accuracy.

The discussion above was focussed on data wires. However, from a circuit perspective, the global clock wires are also treated similarly and lends to similar characterization and timing analysis. This allows HASTA to know the latency and skew in the clock propagation. Additionally, if the PLL specifications are known, HASTA can be supplied with the jitter information to factor the jitter in its analysis. In essence, HASTA can deal with timing analysis of global clock tree with the same degree of accuracy as the conventional EDA tools. A method to synthesize and analyse regional clock tree (**RCT**) in synchoros VLSI design is built on similar (not the same) principles and has been rigorously validated against the commercial EDA tools [7].

HASTA, like the conventional STA, also needs to analyse the flop-to-flop timing in GNoC to check for the following violations: a) setup, b hold, c) combinational delays greater than one clock-period and d) unbuffered clock wires greater than half a clock period of highest intended global clock frequency. Any flop-to-flop timing path is a cascade of the 900 types of segments discussed above.

Listing 3. GNoC synthesis algorithm

```
1.  cost_of_best_solution = infinity;
2.  current_solution = "ALL WIRES"
3.  LOOP
4.  | IF is_valid(current_solution) THEN
5.  |   break;
6.  | IF not (WIRE_BLOCKS ∈ current_solution) THEN
7.  | | current_solution =
8.              remove_buffers(current_solution);
9.  | | current_solution =
10.             insert_registers(current_solution);
11. | ELSE
12. | | current_solution =
    insert_buffers(current_solution);
13. | END IF
14. END LOOP
```

The flop-to-flop timing path analysis needs to distinguish between two cases: forward and backward paths. In a forward path, the clock and the data wires move in the same direction and the positive skew has the effect of stretching the clock period. The reverse is true for the backward path in which the clock and data move in the opposite direction.

*OCV and PVT*

HASTA, like the conventional STA in the EDA tools needs to factor in the PVT and OCV. This is naturally handled because the timing models that HASTA uses are based on data extracted from STA that derates the calculated value for the worst case and the best case. The two-dimensional table-based timing models discussed in section II.a have two variants, one for the best and the other for the worst case. When checking for the hold violations, the min delay table is used and for the setup violations, the max delay table is used.

In essence, when HASTA declares a GNoC path to be timing clean, it is done with same level of rigour and accuracy as the conventional STA tools. The difference is that HASTA works at higher abstraction of 900 types of pre-characterized segments and the timing of arbitrary GNoC is composable in terms of the timing models of these segments.

### C. *The GNoC synthesis algorithm*

In this section, we present an algorithm for the synthesis of GNoC. The synthesis algorithm is presented in Listing 3.

The lowest cost solution is to realize the GNoC links is to use just the plain wire (`W`) SBs, provided that the max frequency and slew rate constraints are met. If these constraints are not met, the next best option is to use buffered wire (`B`) SBs. The `B` SBs are added in increments of one in successive attempts. These blocks are evenly placed. The first attempt is to place a single `B` SB in the middle of the link. If that does not fulfil the requirement, it is removed (line 7) and two buffers, placed at equidistant, are inserted. Adding buffers will lower latency, reduce the slew rate but increase the power consumption. However, allowing slew rate to deteriorate, even if it is within the technology design rules range, will make slower transitions, and increase the crow-bar current that will increase power consumption.

If the buffered wire SBs fail in meeting the frequency constraint, the synthesis algorithm, turns its attention to using the registered wire (`R`) SBs, to pipeline the GNoC wires. The synthesis algorithm begins by inserting a single register and then each registered segments is synthesised using the plain wires and/or buffered wired SBs as described above. If a single register is insufficient, the synthesis algorithm will increment the registered wire (`R`) SBs successively and as with the `B` blocks, placing them evenly; this of course requires eliminating the previously inserted `R` and `B` blocks.

In essence, the synthesis algorithm attempts to find the lowest cost solution and if that does not meet the constraints, it successively adds more expensive SBs, first `B` and then the `R` SBs.

The synthesis algorithm for GCT, in general, requires dealing with cyclic graph. In our present, proof-of-concept, design framework, we simplify the problem by cutting the loop to create an acyclic graph. The principal constraint on GCT is to use buffered (B) and plain wire (W) SBs, such that the max distance between neighbouring B SBs is less than half the distance of the max clock frequency of GCT. Once

Table 1. Experiment results for the delay calculation of different paths

| Path composition | Length | Predicted delay (ns) | EDA tool delay (ns) | Error |
|---|---|---|---|---|
| RWWBWWR | 7 | 0.326 | 0.324 | 0.62% |
| RWWWWWWBWWWWWWR | 15 | 0.667 | 0.662 | 0.76% |
| RBWWBWWWBWWWWBWWWWBWWWBWWBR | 27 | 1.005 | 0.997 | 0.80% |
| RBWWBWBWBWBWWBWWBWWBWBWBWWBR | 27 | 0.814 | 0.808 | 0.74% |

this is achieved, the slew rate is checked and if required, additional B SBs are added.

## III. EXPERIMENTS AND RESULTS

In this section, we present proof-of-concept experimental results to validate the claims a) that it is possible to create GNoC link from a small set of pre-design GNoC SBs. AND B) the timing is known with post-layout accuracy.

The experimental setup involved using EDA tools to create the timing models and to validate the GNoC designs created by abutment and 40 nm technology library.

Three GNoC wire SBs and one switchbox SB were designed. The switchbox SB is a dummy switchbox with flop interface to mimic an arbitrary switchbox. The characterization of internals of switchbox is also predictable and has been covered by [8]. In this paper, the focus is on building and validation of the GNoC links. Including the delays of a specific switchbox/router will simply add a fixed offset in terms of physical design implementation.

To validate the claim, we created synthetic statements based on the grammar listed in Listing 2. These are shown in Table 3. A GNoC VLSI design was constructed corresponding to the four cases by abutting the SBs that corresponds to the sequence of letters in the GNoC sentence.

The composite VLSI design created by abutment was analysed using the timing models discussed in section II.C and also by the commercial EDA tool's STA engines. This analysis was done for setup checks. As can be seen, the value predicted by the GNoC SB timing models is in pretty good agreement with the values predicted by the EDA tools. Figure 4 shows the results of the timing analysis on the last path of Table 3. The x-axis represents the buffer stage and on the y-axis, arrival times and delay are plotted for the results obtained with our method versus the results obtained by the commercial EDA tool. The figure also shows the arrival time comparison between the HASTA and EDA STA. The difference is less than 1%. The cause for the slightly pessimistic but safe results is that when there is not an exact match with the slew rate entry, we used the closest worse slew rate. Using an interpolation of two nearest slew rates and/or increasing the number of slew rate entries would further reduce the error.

The results for the hold-check are not shown because, it is in principle the same method, it just involves using a different characterization table.

## IV. STATE OF THE ART

The NoC research community has richly addressed the challenge of predictable NoC design with quality of service **(QoS)** guarantees [4][5][6], However, these research minimize the uncertainty at architectural and algorithmic level and do not factor in the physical design when the actual delay in wires become known. The method proposed in this paper complements these research by eliminating the physical design uncertainty.

Bertozzi et al. presented a NoC library called xpipes [10] that consists of *soft* macro-like blocks that can be used to build arbitrary GNoC designs. It is important to note that this library does not handle the clocking of the NoC and that task is left to the physical design. Later, the same group at Univ. Of Ferrara along with Teklatech A/S in Denmark, developed the iNOC approach, where a vertically integrated approach of predictable NoC design was attempted. In this framework, starting with a process graph annotated with average traffic on the edges, a floor planning tool factors in the traffic cost while doing floor planning. However, it still relies on backend synthesis to do the final design. In contrast, the synchoros VLSI design, eliminates the logic and physical synthesis and enables creation of arbitrary GNoCs simply by abutment and to know the costs with post-layout accuracy.

Further, the significance of factoring in physical design during NoC synthesis received boost when a special session was organized in NOCS 2020 by Chris Batten and Mike Taylor on "Physical Design Issues for NoCs". This session did bring into focus the need to keep physical design into focus but did not offer any method to factoring in physical design into NOC synthesis as is proposed in this paper.

Dally et al. in their 2013 paper listed physical composition of NoC based on pre-characterised blocks as one of the required steps in moving towards *21st century EDA tools* [11]. This vision paper highlighted the need for a physical regularity in VLSI that would lower the engineering costs of new designs. The synchoros VLSI design framework has similar goals and goes beyond [11] in that it uses synchoros design objects to not just make physical design process simpler but also empower higher abstraction synthesis as envisaged in the DSE loop in Listing 1 and proto-tool for application level synthesis built on this principles [12]. Moreover, Dally et al. [11] do not talk about composing NoC or its analysis and synthesis in terms of SiLago like blocks.

Commercial EDA tools also provide hierarchical timing analysis (HTA) capability [13]. In this method, each block is characterised and its internal logic removed, as we do for the switch box, and only make its relevant peripheral model available for a hierarchical timing analysis. This method has some resemblance to the method proposed in this paper, However, the crucial difference is that HTA is applied to ad-hoc designs and the peripheral model is extracted for each block in each design. Additionally, the logic and physical synthesis and timing analysis at the higher level is still performed, i.e., ad hoc wires do get created and must be analysed; only the internals of block level designs is eliminated for the purpose of efficiency. In the method proposed in this paper, the timing models are extracted as a onetime effort and no new wires (functional or infrastructural) gets created.

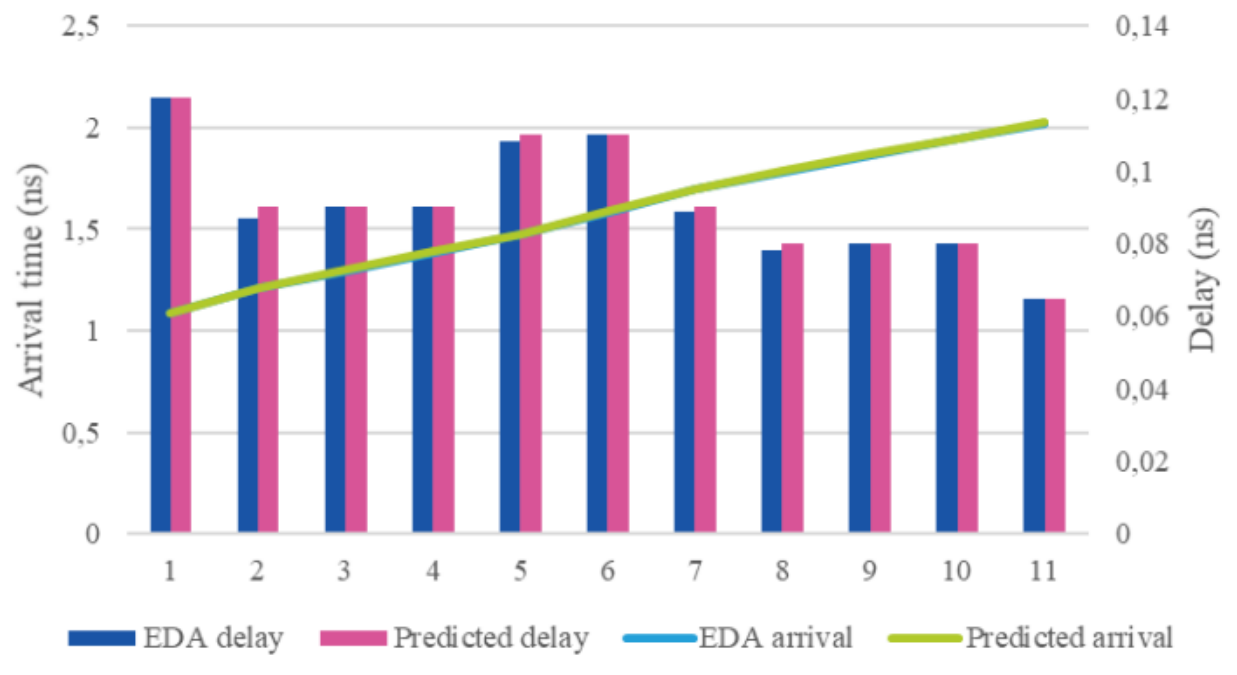

Figure 4. Comparison between the predicted delays by the proposed estimation method and commercial EDA tool

Research effort has also been focused on clocking strategies specifically tailored for NoC designs [14]–[16]. These methods do not address the physical composition of the NoC in terms of hardened SBs to create valid predictable designs.

SiLago's composition by abutment is inspired by the Mead-Conway methodology [17] that later evolved into what became known as silicon-compilation. The Cathedral series of synthesis tools from IMEC were prime examples of this style of design flow, also known as full-custom design flow. There are two key innovations that takes the SiLago design flow beyond previous work: i) SiLago introduces the concept of synchoricity as the basis for making floor-planning and standardize composition by abutment among all SiLago blocks, like Lego bricks. ii) In SiLago, all metal level details are absorbed as part of SiLago blocks. This means that no ad-hoc wires need to be synthesized; not even the infrastructural wires like power-grid, clocks, reset etc. This is fundamental to making synchoros designs predictable with post-layout accuracy and eliminate logic and physical synthesis. In the previous full-custom design flows, these infrastructural wires had to be synthesized anew for each design depending on the floorplan.

## V. Conclusion and Future Work

We have presented a scheme to synthesize arbitrary global NoC from a small set of GNoC SBs. The cost of GNoC is known with post-layout accuracy in linear time. This is envisaged as a critical problem in exploring the design space at chip level, where the interconnect cost is very dominant because wires are not scaling in proportion to logic and storage. The synthesized GNoC is a valid VLSI design that is timing clean.

The experimental framework presented in this paper is proof-of-concept. It is being extended to also predict the energy cost. To some extent, this has been presented in [8] for other SiLago blocks. Another critical enhancement that we are working on is to make the synchoros GNoC designs generated by abutment to be also DRC clean. Rather than find and fix design rule violations in every design and design iteration, it is more productive to factor them into SiLago blocks, as a onetime effort, so that the emerging design is not just timing but also DRC clean.

## VI. Acknowledgements

Supported by CREST II project funded by Vinnova, Sweden.